\documentclass[multphys,vecphys]{svmult}

\usepackage{makeidx}         
\usepackage{graphicx}        
\usepackage{multicol}        
\usepackage[bottom]{footmisc}

\makeindex             

\begin{document}

\title*{Halo Shapes, Dynamics and Environment}
\titlerunning{Halo Shapes, Dynamics and Environment} 
\author{Manolis Plionis\inst{1}, Cinthia Ragone\inst{2} \and
Spyros Basilakos\inst{3}}
\institute{Institute of Astronomy \& Astrophysics, NOA, Athens, Greece \&
  INAOE, Puebla, M\'exico
\and IATE, Observatorio Astron\'omico, C\'ordova, Argentina 
\and Kapteyn Institute, University of Groningen, the Netherlands}
%
%
\maketitle

\begin{abstract}
In the hierarchical structure formation model cosmic 
halos are supposed to form by accretion of
smaller units along anisotropic direction, defined by large-scale
filamentary structures.
After the epoch of mass aggregation (which depend on the
cosmological model), violent relaxation processes will tend to
alter the halo phase-space configuration 
producing quasi-spherical halos with a relatively smooth density profiles.

Here we attempt to investigate the relation between halos shapes,
their environment and their dynamical state. To this end
we have run a large ($L=500 \; h^{-1}$ Mpc,
$N_{p}=512^3$ particles) N-body simulation of a 
flat low-density cold dark matter model with
a matter density $\Omega_{\rm m}=1-\Omega_{\Lambda}=0.3$, Hubble 
constant $H_{\circ}=70$ km s$^{-1}$ Mpc$^{-1}$ and a normalization parameter
of $\sigma_{8}=0.9$. The particle mass is
$m_{\rm p}\ge 7.7\times 10^{10}\,h^{-1}\,M_{\odot}$ comparable to the
mass of one single galaxy. The halos are defined using a friends-of-friend 
algorithm with a linking
length given by $l=0.17\bar{\nu}$ where $\bar{\nu}$ 
is the mean density. This linking length corresponds to an overdensity 
$\rho/\rho_{\rm mean}\simeq 200$ at the present epoch ($z=0$) and 
the total number of halos with more than 130
particles ($M>3 \times 10^{13} \,h^{-1}\,M_{\odot}$) is 57524.
\end{abstract}

\section{Halo Shapes and Environment}
Halo shapes are defined by diagonalizing  the
moments of inertia from which we derive the eigenvalues 
and eigenvectors of the best ellipsoidal halo fit. 
The principal axes $a, b, c$ are related to the square root
of the eigenvalues such that $a>b>c$.
Our results, in agreement with other
studies, indicate that although halos are triaxial they are significantly more
prolate than oblate. This is quantified by using the
so called triaxiality index \cite{Franx} defined as 
$T=(a^2-b^2)/(a^2-c^2)$, which has limiting values of $T=1$ and $T=0$
for the case of a pure prolate and oblate spheroid, respectively. 
Our results show that the fraction of halos with
pronounced prolatness (ie., large $T$s) is significantly higher than that of
oblate-like halos. Overall we obtain from our
simulated halos that $\langle T \rangle\simeq 0.73$.

An interesting question, especially for observational studies, is
whether the 3D halo shape distribution can
be inferred from the projected (2D) shapes. This is an inversion
problem for which, under the assumption of random halo
orientation with respect to the line of sight and purely oblate or
prolate spheroidal halos, there is a unique relation 
between the projected and intrinsic axial ratio distributions. This is
described by a set of integral equations, first investigated by \cite{Hub26}
and given by \cite{Sand70}:
\begin{equation}
\hat{f}(q)=\frac{1}{q^{2}}\int_{0}^{q}\frac{\beta^{2}\hat{N}_{p}(\beta) 
{\rm d}\beta}
{(1-q^{2})^{1/2}(q^{2}-\beta^{2})^{1/2}} \;, \;
\hat{f}(q)=q\int_{0}^{q}\frac{\hat{N}_{\circ}(\beta) {\rm d}\beta}
{(1-q^{2})^{1/2}(q^{2}-\beta^{2})^{1/2}}
\end{equation}
with $\beta$ representing the intrinsic axial ratio while 
$\hat N_o(\beta)$ and $\hat N_p(\beta)$ the intrinsic prolate
and oblate axial ratio distributions, respectively. 
The continuous function $f(q)$ is derived from the discrete axial
ratios frequency distribution using the so-called kernel
estimators (for details see \cite{Ryden} and references therein).
Inverting then the above equations
gives us the distribution of true axial ratios as a function of $f(q)$
(eg. \cite{Fall}). Nevertheless, if halos are a mixture of
the two spheroidal populations or they have triaxial configurations
there is no unique inversion \cite{Pl91}. However, all may
not be lost and although the exact  shape distribution may not be
recovered accurately one could possibly infer whether the 3D halo
shapes are predominantly more prolate or oblate-like. Let us see this in
more detail using our simulated intrinsically triaxial ellipsoidal
halos which are however dominated by prolate-like shapes ($\langle T
\rangle \simeq 0.73$).
The important point here is that in order for the inverted distribution
to be physically meaningful it should be positive for all
$\beta$'s. Negative values indicate that the assumed model for the
intrinsic halo shape is unacceptable.
\begin{figure}
\centering
\includegraphics[height=7.8cm]{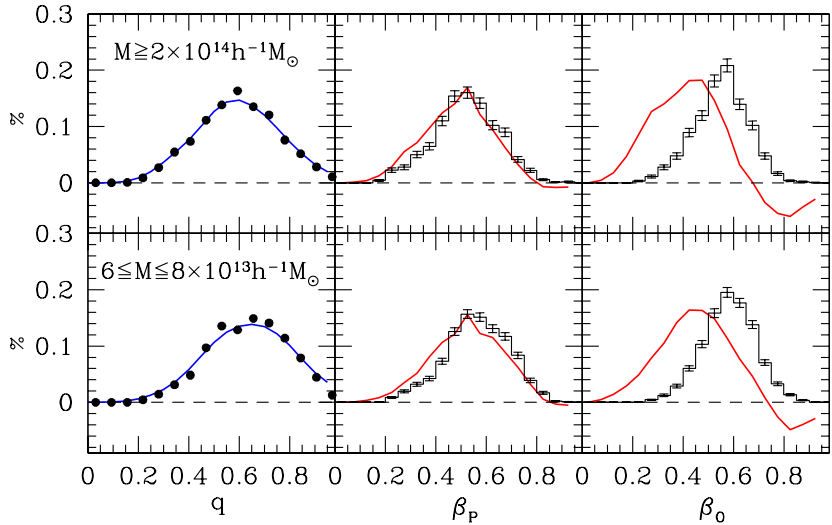}
\caption{{\em Left Panel:} The projected axial ratio distributions for
  two different halo mass thresholds with the nonparametric kernel
  estimator fit (solid line). {\em Central
  Panel:} The inverted intrinsic halo axial ratio
  distribution (continuous line) for the prolate model together with
distribution of ``average'' prolate spheroidal fit to the 3D halos. 
{\em Right Panel:} the corresponding distributions for the oblate case.}
\end{figure}

In Fig.1 (left panel) we present the discrete and continuous - $f(q)$ -
distributions of the projected in 2D axial ratios for halos of 
two mass ranges (indicated in the plot). In the middle and right
panels we present the inverted 3D axial ratio distributions 
(continuous lines) for the prolate and oblate models, respectively. 
It is evident that the inverted oblate-model distribution 
has many negative values which is an important indication
that this model is unacceptable, while the opposite is true for the
prolate-model distribution. Furthermore, we plot as histograms
the intrinsic axial ratio distribution of the ``average'' prolate or
oblate spheroidal fit to the 3D halo. 
These fits are realized by estimating the corresponding axial ratios
by $\beta_{\rm P}=(b+c)/2a$ and $\beta_{\rm O}=2c/(a+b)$.
It is evident that the purely
oblate model fails miserably to even come close to the inverted
distribution while the prolate model fits relatively well the
corresponding inverted 3D prolate-model distribution. This agrees with
the higher prolatness of the 3D halo shapes. We therefore
conclude that applying the previously discussed
inversion method to observational data (eg. \cite{Pl04}, \cite{Paz})
we can infer, even in the event of triaxial
ellipsoidal halo shapes, the dominance of prolate or oblate-like 3D
shapes.

Another interesting fact, shown in recent high-resolution simulations
of the concordance model, is
the correlation between halo mass and halo flattening,
with more massive halos being flatter (eg. \cite{All05},
\cite{Kasun}, \cite{Jing02}). This is counter-intuitive in
the sense that the massive halos should collapse faster
than lower mass halos of roughly 
the same formation age and thus they should have more time to
dynamically evolve and virialize, a process that should reduce
their initial ellipsoidal configuration. However, halo formation
ages vary and secondary infall, which if important it could affect the 
halo outskirts, can produce such elongated halo geometry.
However, even in such a case
the inner parts of the most massive halos should be virialized and
thus nearly spherical,
which however does not seem to be the case (eg. \cite{All05} their
figure 7). Note that these results are based on
dissipationless simulations while
baryonic dissipation has been shown to affect halo shapes, producing
significantly rounder halos (eg. \cite{Kaz04}).
Also one may expect that halos in low-density regions, where tidal
effects, accretion and merging are minimal to be less elongated, 
as indeed has been found (eg. \cite{Avila}).
 
\begin{figure}
\centering
\includegraphics[height=5.8cm]{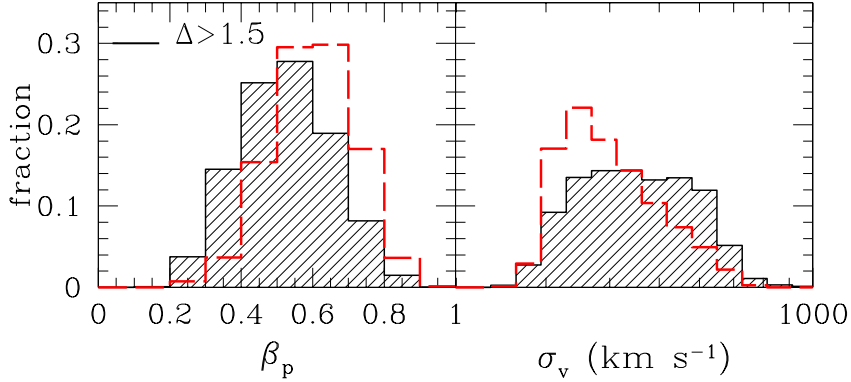}
\caption{Normalized axial ratio and velocity dispersion distributions. 
Hatched histograms represent halos with significant substructure while the
dashed-line histograms represent the overall halo population.}
\end{figure}

\section{Halo Dynamics \& Environment}
In order to define the dynamical state of a halo we use the 
$\Delta$-deviation substructure statistic \cite{Dress}, 
which looks for deviations of the local 
velocity mean and dispersion around their overall
halo values. For each halo particle we find its $N_{nn}$
nearest neighbors from which we calculate their local velocity mean
and dispersion which we then compare with their overall halo values.
The local deviations are defined by:
$\delta_i^2=\frac{N_{nn}}{\sigma_v^2} \left[ (\bar{v}_{local}-\bar{v})^2
  + (\sigma_{v,local}-\sigma_v)^2\right]
$
while the individual halo $\Delta$-statistic is given by the sum 
$\Delta=\sum \delta_i$. It has been found that a robust measure of the 
substructure index is given when using $N_{nn}=25$ \cite{Knebe00}, 
which is the value that we use.
In figure 2 we show the velocity dispersion (right panel) and 
$\beta_P$ axial ratio (left panel) distributions for halos
in the mass range $3 \times 10^{13} < M < 8\times 10^{13} \; h^{-1} \;
M_{\odot}$.
Those that have a substructure index $\Delta >1.5$ are shown as
hatched histograms while the overall halo distributions as dashed-line
histograms. {\em It is evident that dynamically young halos (ie., those
with significant substructure) are more elongated and have a higher
velocity dispersion than the overall halo population.}

Furthermore, we investigate the correlation between halo velocity
dispersion and halo mass. 
From our halo identification procedure we expect that most
halos should be nearly virialized and thus their velocity
dispersion should be strongly correlated with their mass via the 
virial theorem. Indeed, there is a strong
correlation between the velocity dispersion and the halo mass, 
measured by summing the member DM particles masses, with a Spearman
correlation coefficient of $R\sim 0.9$. 
\begin{figure}
\centering
\includegraphics[height=6.3cm]{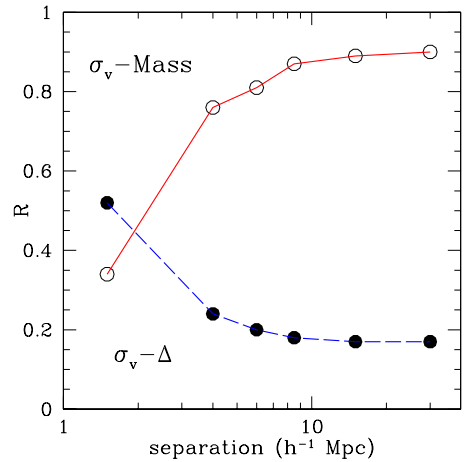}
\caption{Spearman correlation coefficient for the $\sigma_v$ vs Mass and 
$\sigma_v$ vs $\Delta$ correlations as a function of host-halo distance.}
\end{figure}
We find however, that this strong
correlation breaks down in the vicinity of large halo hosts. In figure
3 we show with open points the correlation coefficient between halo $\sigma_v$ and Mass
as a function of distance to their nearest massive ($M>2\times 10^{14}
\;h^{-1}\; M_{\odot}$) host neighbor.  The monotonic
drop of the correlation coefficient with decreasing halo-host
separation is evident. We also
find that there is a significant although weaker correlation between $\sigma_v$ and
the substructure index ($\Delta$), which increases as the
halo-host distance decreases. This probably implies that halos near large hosts
are either disrupted due to the presence of strong tidal fields or
more probably that the excess density of sub-halos near hosts induce 
strong inter-halo gravitational interactions and merging.
These results have important consequences for observational studies
and put important limits on the validity of using the virial theorem to
estimate group or cluster masses in the vicinity of massive clusters.



\printindex

\begin{thebibliography}{99.}
\bibitem{All05} Allgood et al., ApJ, {\em in press}, {\tt astro-ph/0508497} 
\bibitem{Avila} Avila-Reese, V., et al., 
ApJ, \textbf{634}, 51 (2005)
\bibitem{Dress}Dressler, A. \& Shectman, S.A., AJ, \textbf{95}, 985
  (1988)
\bibitem{Fall}Fall, M. \& Frenk, C. S., ApJ, \textbf{88}, 1626
  (1983)
\bibitem{Franx}Franx, M., Illingworth, G., de Zeeuw, T., ApJ,
  \textbf{383}, 112 (1991)
\bibitem{Hub26}Hubble, E.P., ApJ, \textbf{64}, 321 (1926)
\bibitem{Jing02}Jing, Y.P. \& Suto, Y., ApJ, \textbf{529}, L69 (2002)
\bibitem{Kasun} Kasun, S.F. \& Evrard, A.E., ApJ, \textbf{629}, 781 (2005)
\bibitem{Kaz04}  Kazantzidis, S. et al., ApJ, \textbf{611}, L73 (2004)
\bibitem{Knebe00} Knebe, A., M\"uller, V., A\&A, \textbf{354}, 761 (2000)
\bibitem{Paz} Paz, D.J., Lambas, D.G., Padilla, N., Merchan, M.,
  MNRAS, {\em in press} (2006)
\bibitem{Pl91}Plionis M., Barrow J.D., Frenk, C.S., MNRAS, \textbf{249},
662 (1991)
\bibitem{Pl04}Plionis M., Basilakos, S., Tovmassian, H., MNRAS, \textbf{352}, 1323 (2004)
\bibitem{Ryden} Ryden, B.S., ApJ, \textbf{461}, 146 (1996)
\bibitem{Sand70} Sandage, A., Freeman, K.C. \& Stokes, N.R., ApJ,
  \textbf{160}, 831 (1970)



\end{thebibliography}
\end{document}